\documentclass[aps,pra,superscriptaddress,twocolumn,showpacs,titlepage=false,floatfix]{revtex4-1}
\usepackage{amsmath, amsthm, amssymb, graphicx, txfonts}

\usepackage{listings}
\usepackage{color}
\usepackage{algorithm}
\usepackage{algpseudocode}
\usepackage[english]{babel}

\begin{document}

\newcommand{\reals}{\mathbb{R}}
\newcommand{\complex}{\mathbb{C}}
\newcommand{\expect}[1]{\ensuremath{\left\langle#1\right\rangle}}
\newcommand{\ket}[1]{\ensuremath{\left|#1\right\rangle}}
\newcommand{\bra}[1]{\ensuremath{\left\langle#1\right|}}
\newcommand{\braket}[2]{\ensuremath{\left\langle#1|#2\right\rangle}}
\newcommand{\ketbra}[2]{\ket{#1}\!\!\bra{#2}}
\newcommand{\braopket}[3]{\ensuremath{\bra{#1}#2\ket{#3}}}
\newcommand{\proj}[1]{\ketbra{#1}{#1}}
\newcommand{\rrangle}{\rangle\!\rangle} \newcommand{\llangle}{\langle\!\langle}
\newcommand{\sket}[1]{\ensuremath{\left|#1\right\rrangle}}
\newcommand{\sbra}[1]{\ensuremath{\left\llangle#1\right|}}
\newcommand{\sbraket}[2]{\ensuremath{\left\llangle#1|#2\right\rrangle}}
\newcommand{\sketbra}[2]{\sket{#1}\!\!\sbra{#2}}
\newcommand{\sbraopket}[3]{\ensuremath{\sbra{#1}#2\sket{#3}}}
\newcommand{\sproj}[1]{\sketbra{#1}{#1}}
\def\Id{1\!\mathrm{l}}
\newcommand{\Tr}{\mathrm{Tr}}
\newcommand{\cM}{\mathcal{M}}
\newcommand{\cP}{\mathcal{P}}
\newcommand{\cG}{\mathcal{G}}
\newcommand{\cD}{\mathcal{D}}
\newcommand{\cE}{\mathcal{E}}
\newcommand{\cL}{\mathcal{L}}
\newcommand{\cU}{\mathcal{U}}
\newcommand{\cH}{\mathcal{H}}
\newcommand{\cC}{\mathcal{C}}

\newcommand{\bvec}[1]{\ensuremath{\mathbf{#1}}}
\newcommand{\todo}[1]{\textcolor{red}{#1}}

\renewcommand{\floatpagefraction}{.95}

\title{Efficient flexible characterization of quantum processors with nested error models}
\begin{abstract}
We present a simple and powerful technique for finding a good error model for a quantum processor.  The technique iteratively tests a nested sequence of models against data obtained from the processor, and keeps track of the best-fit model and its wildcard error (a quantification of the unmodeled error) at each step.  Each best-fit model, along with a quantification of its unmodeled error, constitute a characterization of the processor.  We explain how quantum processor models can be compared with experimental data and to each other.  We demonstrate the technique by using it to characterize a simulated noisy 2-qubit processor.
\end{abstract}

\author{Erik Nielsen}
\author{Kenneth Rudinger}
\author{Timothy Proctor}
\author{Kevin Young}
\author{Robin Blume-Kohout}
\affiliation{Quantum Performance Laboratory, Sandia National Laboratories, Albuquerque, NM 87185 and Livermore, CA 94550}

\date{\today}
\maketitle

\section{Introduction}
A quantum processor consists of a collection of effective 2-level physical systems called qubits, and a system that regulates and controls these qubits and their environment \cite{DiVincenzo2000-iw, Ladd2010}.  In order for the processor to work properly, the control system must maintain the coherence of the qubits' collective quantum state while performing very specific manipulations of that state.  In real quantum processors these \emph{quantum logic operations} act imperfectly\cite{Preskill2018NISQ}.  This limits the processor's computing power and utility.  Understanding these imperfections is critical to improving future hardware and advancing the state of the art \cite{Eisert2020}.

There exist a variety of \emph{QCVV} (quantum characterization, verification, and validation) protocols that aspire to identify and quantify various deviations from ideal processor behavior.  Most (arguably all) of these techniques rely on some \emph{model}, implicit or explicit, for the those deviations.  Some use comprehensive models that describe the fine-grained behavior of every logic operation, and are intended to predict the outcome probabilities of arbitrary quantum circuits \cite{MerkelPRA13, Greenbaum15, Blume-Kohout2017-kn, Sarovar2019-xc, Proctor2020-iz, matteo2020operational, Nielsen2020-lu}.  Other QCVV techniques use simpler models for coarse-grained observable properties -- e.g., binary success/failure probabilities, or specific circuits only, or averages over circuit ensembles \cite{Cross2019-ku, Boixo2018-kp, Magesan2011-hc, Knill2008-jf, Proctor2019-wp, EmersonScience2007, Magesan2012-dz,Carignan-Dugas2015-bi, proctor2020measuring}.  ``Good'' models of either type -- i.e., ones that accurately fit the data -- can be used to identify noise processes and error mechanisms in the quantum hardware, to extrapolate the behavior of existing devices, and to predict the behavior of future hardware.  

Model \emph{complexity} presents a fundamental trade-off.  Complex models with many adjustable parameters, such those used in gate set tomography (GST) \cite{Blume-Kohout2017-kn, nielsen2020gate}, often provide greater predictive power, more robustness to unanticipated phenomena, and more insight into error sources.  But these virtues come at a cost.  Complex models are computationally harder to evaluate and trickier to interpret, and fitting their many parameters demands more experimental data.  Simpler models, such as the 3-parameter model used by randomized benchmarking (RB) \cite{Magesan2011-hc, Knill2008-jf, Proctor2019-wp, EmersonScience2007}, are much easier to construct and interpret, and can be more easily scaled to larger number of qubits -- but they are often less predictive, and provide less insight into underlying physical mechanisms.  So choosing a QCVV protocol (and its associated model of errors) at the beginning of an experiment can be a momentous choice -- a ``too simple'' model won't capture all the errors, while a ``too complicated'' model will demand excessive resources.


In this work, we introduce a new paradigm.  Instead of choosing a protocol and a model in advance, we dynamically explore a \emph{range} of models and experimental designs to find one that explains the processor's behavior parsimoniously.  This approach, and the specific technique we deploy, are motivated by two key take-aways from our experience characterizing experimental processors: (1) There's rarely such a thing as ``the right'' noise model for an experimental processor; and (2) it's critical to balance the model richness needed to describe observed data against the simplicity needed to facilitate useful interpretation.

Our basic methodology is to construct a set of \emph{nested} candidate models, arrange them in a sequence, then iteratively fit them against data and use statistical tests to determine whether to proceed further (adding more data), or try a bigger model.  Testing a statistical model is a well-researched task. There are powerful statistical methods for quantifying when a model is consistent with a set of data, and when a particular model should be preferred over another.  We apply these methods -- and some novel ones that we developed recently -- to the task of finding empirical models for errors in quantum processors.

Section \ref{sec:multimodel} provides additional motivation for our proposed method, and summarizes it at a high level.  Section \ref{sec:testing} introduces important technical background and definitions, and describes how statistical model testing can be applied within the context of quantum characterization.  Then, in Section \ref{sec:method}, we present our method completely and discuss its properties.  Finally, in Section \ref{sec:example} we demonstrate our protocol by applying it to the simple case of a simulated 2-qubit quantum processor.  We use the open source \texttt{pyGSTi} software package \cite{pygsti, Nielsen2020-lu} to perform the numerical analysis.

\section{Characterization using multiple models\label{sec:multimodel}}
Characterizing a quantum processor typically involves choosing a method (e.g.,~RB or GST), based on a single model, and running it.  Such methods use models with varying strengths and sizes, but in the end only a single model is ever utilized and we must accept the strengths and weaknesses inherent in it.

Standard gate set tomography \cite{Blume-Kohout2017-kn, nielsen2020gate}, for example, uses a large model where each gate is an arbitrary CPTP map.  The best-fit of this model is compared with the data, with the hope that it will describe most if not all of the data.  If it does, the large best-fit model must still be analyzed to extract meaningful simple metrics that describe the errors in an intuitive way.  This approach can be inefficient because the GST model has more parameters than are usually needed.  It allows for the possibility of many errors that either 1) don't occur in the device or 2) aren't intuitive or aren't related to hardware adjustments that could improve the device.  A large model can also make the analysis (e.g. fitting the model) time consuming and the interpretation of the result opaque.  Finally, large models require a proportionately large amount of data to unambiguously estimate all of their parameters, which demands more experimental resources.  Standard 2-qubit GST requires thousands or tens of thousands of circuits.  In many cases, most of these circuits are unnecessary because they probe errors that aren't present in the device.

Randomized benchmarks \cite{Magesan2011-hc, Knill2008-jf, Proctor2019-wp, EmersonScience2007, Magesan2012-dz, Carignan-Dugas2015-bi} suffer from complementary ailments. For example, standard RB's model contains a single gate error rate for an ``average Clifford gate'' \cite{Magesan2011-hc}. This model is not intended to predict the outcomes of arbitrary circuits, and it can be difficult to generalize this error rate into meaningful statements about processor performance.

\begin{figure}
  \begin{center}
    \includegraphics[width=0.8\columnwidth]{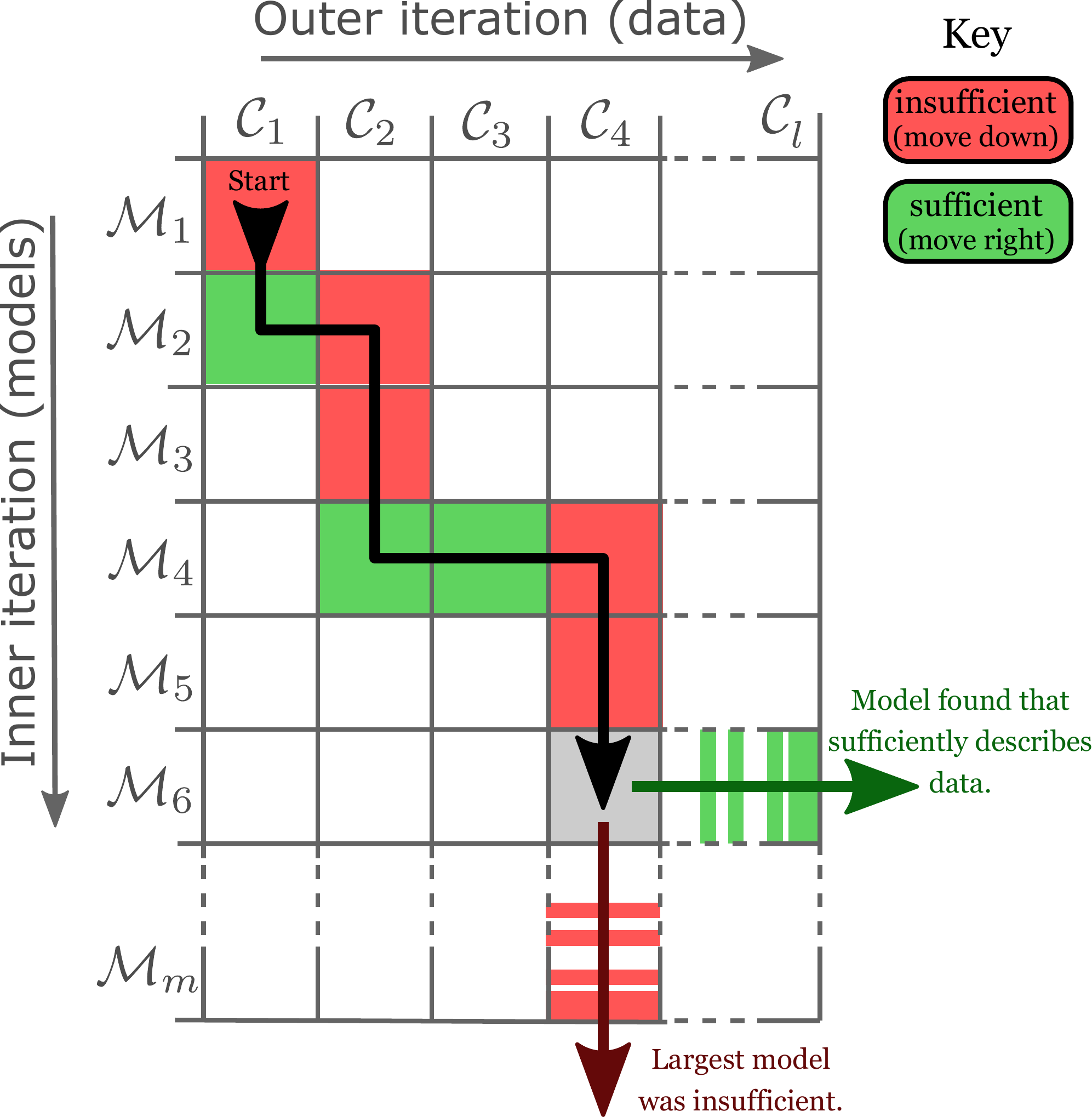}
  \end{center}
  \caption{\textbf{The multi-model characterization method.} A graphical depiction of Algorithm \ref{lst:model_testing_method}.  Each iteration of the algorithm either advances to a larger dataset (when the current model is accepted) or advances to a larger model (when the current model is rejected).  These advances are depicted by downward and rightward moves, respectively, on a 2D grid indexed by the circuit (data) sets and available models.  Green boxes indicate where the current model (row) sufficiently describes the current data (column), prompting a rightward move.  Red boxes indicate where the model was insufficient, prompting a downward move.  The black line and arrow track the path of the algorithm in time.  Green and red arrows show the two possible ways the algorithm ends: either a sufficient model is found for the largest dataset or the available models are exhausted.\label{fig:method_schematic}}
\end{figure}

Solving these problems demands adaptive methods that integrate multiple models during the characterization process.  In this article we present one such multi-model approach, and show how it offers distinct advantages over single-model approaches.  Our procedure takes as input a sequence of nested models, ordered from smallest to largest, and a sequence of \emph{experiment designs} -- lists of circuits that define an experiment that could be performed on a quantum processor.  Beginning with the smallest model and simplest experiment design, we compare our current model to the data from the current experiment design and decide whether the model sufficiently captures the data.  If it does, we move to the next experiment design, to perform a more strenuous test of the model.  If it does not, we move to the next larger model, in hopes that it \emph{will} capture the data.  This process is depicted in Fig.~\ref{fig:method_schematic}.  Instead of fitting a large model to a large amount of data at the outset, and potentially finding that many of its error-rates are zero, we begin with a small model and dataset and test whether anything more is needed to describe the data.  We only move to a larger model if the smaller model is deemed insufficient.

In the next two sections, we make this idea more precise and concrete.  Section \ref{sec:method} restates the procedure of Fig.~\ref{fig:method_schematic} in greater detail and as pseudocode.  In the intervening Section \ref{sec:testing}, we introduce necessary background such as defining what nested models are and identifying metrics that we can use to decide whether a model ``sufficiently describes'' a set of data.  



\section{Model Testing\label{sec:testing}}
Statistical models of logic operations appear naturally when characterizing a quantum processor.  In this section we explain how statistical models relate to models of quantum processors, and how tools from statistics can be used to test and select among them.  After some preliminary definitions we describe how a processor model can be compared with data from an actual quantum processor, and how different models can be compared with each other.

\subsection{Datasets}
A \emph{quantum circuit} describes a sequence of quantum gates (quantum logic operations) on a fixed number of qubits.  All the quantum circuits we consider in this work begin with a state preparation on, and end with the measurement of, all the qubits in the system.  An ordered list of distinct quantum circuits, which we represent using script $\cC$, along with an integer sample count, $N$, specifies an experiment to be performed on a quantum processor.  We also call $\cC$ an \emph{experiment design}.  The outcomes obtained from executing each circuit $N$ times \footnote{The number of repetitions may vary from circuit to circuit, but we do not consider this generality here.} form a \emph{dataset}, which we denote using $\cD = \cD(\cC, N)$.  In this work, we only use the histogram of outcomes for each circuit -- the time-ordering of the outcomes is discarded.  Extensions to time dependent models \cite{Proctor2020-iz, Bennink2019-om} that require time series data, are left for future work. 

\subsection{Models}
A model for a quantum processor is a mathematical object that can predict the outcome of any quantum circuit that is run on the processor.  Since quantum mechanics is probabilistic, this prediction takes the form of a probability distribution over the possible circuit outcomes.  A \emph{quantum processor model} $\cM$ is therefore a parameterized set
\begin{equation}
    \cM = \left\{ \mathrm{M}_{\vec{\theta}}\,:\,\vec{\theta}\in \Theta \right\}
\end{equation}
 of functions $\mathrm{M}_{\vec{\theta}}$ that each map circuits to outcome probability distributions.  We also refer to $\mathrm{M}_{\vec{\theta}}$ as a model, since it is simply a quantum processor model without parameters.  $\Theta$ is $\cM$'s parameter space.  It's dimension, $k$, is the model's \emph{number of parameters}.  When a processor model is combined with an experiment design $\cC$, a \emph{statistical model} -- a parameterized probability distribution \cite{StatisticalModelsBook} -- results.  The value of the statistical model $\cM(\cC)$ at $\vec{\theta}$ is simply the product of the circuit probability distributions $\prod_{c \in \cC} \mathrm{M}_{\vec{\theta}}(c)$.  Thus, $\cM(\cC)$ has $k$ parameters and, at every $\vec{\theta}$, predicts an outcome probability distribution for each circuit in $\cC$.  When the circuits are clear from the context, we will omit them and simply use $\cM$ to denote the statistical model $\cM(\cC)$.  

One way of specifying a model is by associating process matrices with each of a processor's available operations.  By multiplying and contracting process matrices, such a model can be used to predict the probabilities of any circuit and thus for the circuits in $\cC$.  A depolarizing noise model, where the same $n$-qubit depolarizing channel is applied after each gate, is a specific example of a 1-parameter quantum processor model.


\subsection{Comparing a model to a dataset}
A well-established way to quantify how well a model fits a set of data is the log-likelihood statistic.  It is defined between a model $\mathrm{M}$ and dataset $\cD$ as the probability of $\mathrm{M}$ given $\cD$.  If $c$ indexes each circuit for which $\cD$ contains data, and $\beta_c$ the allowed outcomes of $c$, then the log-likelihood is given by
\begin{equation}
\log\cL(\mathrm{M}, \cD) = \sum_{c,\,\beta_c} N_c\, f_{c,\,\beta_c} \log(p_{c,\,\beta_c}),\label{eq:logl}
\end{equation}
where $N_c$ is the number of times circuit $c$ is repeated, $f_{c,\beta_c}$ is the frequency (fraction of total counts) with which outcome $\beta_c$ is observed after running $c$, and $p_{c,\beta_c}$ is the corresponding probability predicted by $\mathrm{M}$.  The log-likelihood is well-justified on a number of counts: the inverse of its Hessian is the Fischer information\cite{Bradley1978fisher} and by definition it quantifies the probability that the model produced the set of data.  Intuitively, it quantifies how surprising it would be for the model to have generated the data.  We define a parameterized model's log-likelihood as the maximum $\log\cL$ over its parameter space, i.e.,
\begin{equation}
  \log\cL(\cM, \cD) = \max_{\vec{\theta} \in \Theta} \log\cL\left(\mathrm{M}_{\vec{\theta}}, \cD \right).
\end{equation}

If a model's predictions exactly match the observed frequencies, then the maximum possible likelihood is reached.  The value for which this occurs is not a universal quantity, and it is a well known fact that the value of $\log\cL$ is only meaningful in a relative sense.  When $\log\cL$ is given by Eq.~\ref{eq:logl}, then this maximum,
\begin{equation}
\log\cL_{\max}(\cD) = \sum_{c,\,\beta_c} N_c\, f_{c,\,\beta_c} \log(f_{c,\,\beta_c}),\label{eq:logl_max}
\end{equation}
is clearly dependent on the observed data.  Typically, a model does not predict the observed frequencies exactly, and we need to determine the quality of ``goodness'' of the fit based on the obtained value of $\log\cL$. 

Model $\cM$ is said to be \emph{valid} relative to $\cD$ if it contains the (non-parameterized) model $\bar{\mathrm{M}}$ that generated $\cD$ -- or a map indistinguishable from $\bar{\mathrm{M}}$.  The \emph{maximal model} for $\cD$, constructed to have one parameter for every independent observable probability in $\cD$, fits $\cD$ perfectly, achieves $\log\cL_{max}(\cD)$, and is valid, as there is no data that can falsify it.  In general, determining a model's validity -- a proxy for its ``goodness'' -- requires comparing it to a valid model.  This makes maximal models particularly important points of reference.

\subsection{Measuring goodness of fit}
Two fundamental quantities enter into the perceived ``goodness'' of a model's fit to a dataset: the $\log\cL$ between model and data, and the number of parameters, $k$, of the model.  When a model is given more parameters, it is able to fit any given set of data at least as well or better, and thus $k$ and $\log\cL$ will trade-off with each other.  Because, in this trade-off, we consider adding or subtracting parameters from a model, the concept of \emph{nested} models is relevant.  We say that $\cM_A$ is nested within $\cM_B$, and write $\cM_A \subset \cM_B$, when $\cM_A$ constitutes a subset (within parameter space) of $\cM_B$.

As we look at ways of measuring a model's goodness of fit, two underlying truths are helpful to keep in mind. First, a model's $\log\cL$ and $k$ are independently important for assessing its fit.  No single number can capture all the information contained in both.  Secondly, we can only assess a model's fit \emph{relative} to that of another model.  This somewhat annoying fact is a consequence of the relative nature of $\log\cL$ discussed above.  There is a sense in which a model's fit cannot be declared ``good'' with complete objectivity.  Thankfully, maximal models provide nearly objective reference points to assess the goodness of other models' fits.

There are, more or less, two different ways to go about quantifying the goodness of a fit.  The first way quantifies the amount of \emph{evidence} of unmodeled effects.  The more conclusive the evidence is against a model's being able to describe all the data, the worse the fit is deemed.  The second way quantifies the \emph{size} of the unmodeled effects.  Here, a model's fit becomes worse as larger and larger corrections are needed to explain the data after starting at the model's predictions.  We look at each approach in turn.

\begin{figure}
  \begin{center}
    \includegraphics[width=0.9\columnwidth]{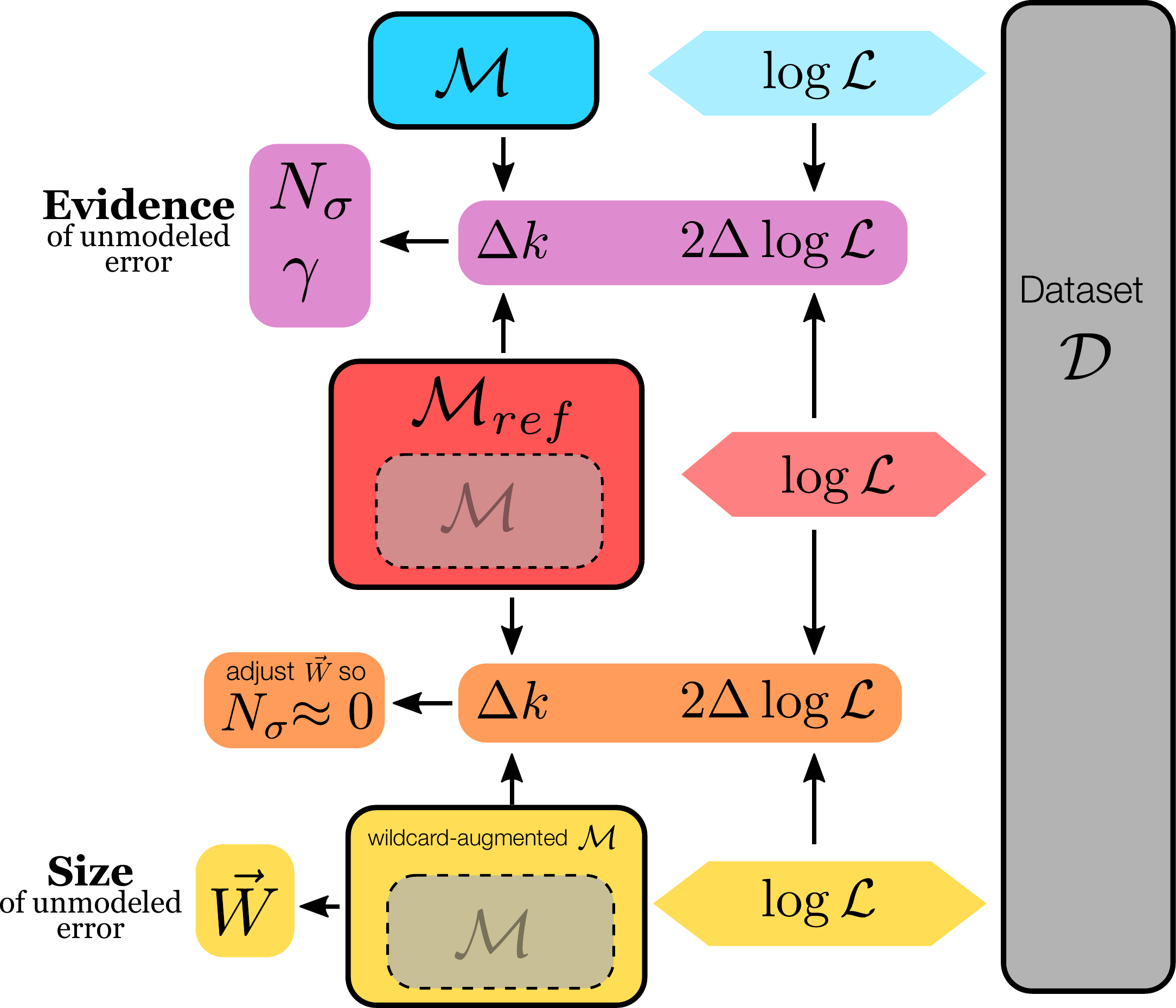}
  \end{center}
  \caption{\textbf{Computing goodness of fit between model and data.}  The fit of model $\cM$ is compared to that of the (valid) reference model $\cM_{ref}$.  Three models enter into the calculation: $\cM$, $\cM_{ref}$ and the wildcard-augmented $\cM_{\vec{W}}$, which predicts balls of probability distributions based on its wildcard error rates $\vec{W}$ (see text).  First, the log-likelihood between each model and a common dataset $\cD$ is computed.  Taking differences of $\log\cL$ values and model parameter counts produces $\Delta k$ and $2\Delta\log\cL$, which in turn are used to compute $N_\sigma$ and $\gamma$ (Eqs.~\ref{eq:Nsigma} and \ref{eq:gamma}).  Between $\cM$ and $\cM_{ref}$ these constitute measurements of the \emph{evidence} that $\cM$ is invalid. The smallest $\vec{W}$ is found for which the $\log\cL$ of the wildcard-augmented model $\cM_{\vec{W}}$ meets a pre-specified goodness-of-fit threshold (see text), written as $N_\sigma \approx 0$, where the ``size'' of $\vec{W}$ is its $L_1$-norm. The size of $\vec{W}$ quantifies the \emph{size} of the errors $\cM$ fails to model.\label{fig:comparisons}}
\end{figure}

\subsubsection{Quantifying the evidence of unmodeled effects}
Suppose we want to know how well model $\cM_A$ fits a set of data $\cD$.  Following our remarks above, this goodness-of-fit will need to be relative to some other model, $\cM_B$.  Let the models' parameter counts be $k_A$ and $k_B$, respectively.  In the language of hypothesis testing, $\cM_A$ is our null hypothesis, $\cM_B$ is our alternative hypothesis, and we want to know whether, or with what certainty, we should reject the null hypothesis.  When $\cM_B$ is chosen such that $\cM_A \subset \cM_B$ and to be valid, then Wilks' theorem \cite{wilks1938} can help answer this question.  Wilks' theorem states that twice the difference in the log-likelihoods, $2\Delta\log\cL = 2\left(\log\cL(\cM_B,\cD) - \log\cL(\cM_A,\cD) \right)$, is asymptotically $\chi^2_{\Delta k}$-distributed, where $\Delta k = k_B - k_A$, \emph{when both models are valid}.  The mean ($\Delta k$) and standard deviation ($\sqrt{2\Delta k}$) of the $\chi^2_{\Delta k}$ distribution effectively bestow an origin and unit, respectively, on the relative $\log\cL$ between nested models.  Since we have assumed $\cM_B$ to be valid, the quantity
\begin{equation}
  N_{\sigma}(\cM_A, \cM_B) = \frac{2\Delta\log\cL - \Delta k}{\sqrt{2\Delta k}},\label{eq:Nsigma}
\end{equation}
measures, in units of standard deviations, how certain we are that $\cM_A$ should be rejected, i.e.~how certain we are that $\cM_A$ is invalid.  In other words, $N_\sigma$ quantifies the amount of evidence that $\cM_A$ is invalid.  Informally, it quantifies how surprising it would be to learn that model $\cM_A$ (at any $\vec{\theta}$) generated $\cD$.

A second metric for comparing $\cM_A$ and $\cM_B$ can be derived from the expected trade-off between $\log\cL$ and $k$.  Removing one free parameter will almost surely decrease $2\log\cL$.  Wilks' theorem tell us that removing a useless parameter (one that takes its true value in the smaller model) causes an expected decrease of 1 unit.  So if we remove $\Delta k$ useless parameters, we expect $2\Delta\log\cL \approx \Delta k$.  Removing a useful parameter (one not constrained to its true value by the smaller model) will decrease $2\log\cL$ more.  The Akaike information criterion (AIC) \cite{1100705_AIC} states that under certain idealized assumptions, removing a set of $k$ parameters will yield an estimate with greater predictive accuracy iff $2\Delta\log\cL < 2\Delta k$.  Both Wilks' theorem and the AIC suggest that we can evaluate the ``usefulness'' of a set of $\Delta k$ parameters with a quantity we call the \emph{evidence ratio},
\begin{equation}
  \gamma(\cM_A, \cM_B) = \frac{2\Delta\log\cL}{\Delta k}.\label{eq:gamma}
\end{equation}
The evidence ratio tells us how much more data-fitting power $\cM_B$ has, \emph{per parameter}, than $\cM_A$.  We can use it to apply a variety of rules for choosing between those models -- i.e. model selection -- simply by choosing a threshold $\xi$.  We declare $\cM_A$ superior to $\cM_B$ whenever $\gamma < \xi$.  Wilks' theorem implies that if all the ``extra'' parameters in $\cM_B$ are useless, then we'll observe $\gamma\approx 1$.  So the threshold $\xi=1$ would only select $\cM_A$ when $\cM_B$ provides absolutely no additional model-fitting power (and even then, only 50\% of the time, at random).  Larger values of $\xi$ favor smaller models, selecting $\cM_A$ even when its validity becomes less certain.  The threshold $\xi=2$ implements the AIC rule \cite{1100705_AIC}.  Often, even higher thresholds are desirable for quantum gate characterization, because model simplicity is prized.  Like $N_\sigma$, the evidence ratio quantifies the amount of evidence for unmodeled errors, not their size.

Both $N_\sigma$ and $\gamma$ give a relative comparison of two models' fit to data.  While this is expected, given the relative nature of the likelihood function, we would like to assess a model's fit in an absolute sense.  We are able to do this effectively by fixing, for a given data set, $\cM_B$ to be a \emph{reference model}, $\cM_{ref}$.  The reference model must be suitably large so as to include (as nested models) all the models we wish to test, and it must be valid (a condition for interpreting $N_\sigma$).  The maximal model introduced above has all of these properties, and for the remainder of this work, we take as the reference model the corresponding maximal model.  After this, $N_\sigma$ and $\gamma$ become functions of only $\cM_A$.  We write them as functions of a single model $\cM$,
\begin{eqnarray}
  N_{\sigma}(\cM) & \equiv & N_{\sigma}(\cM, \cM_{ref}) \\
  \gamma(\cM) & \equiv & \gamma(\cM, \cM_{ref}),
\end{eqnarray}
and omit the argument entirely when the model being compared is clear from the context.

\subsubsection{Quantifying the size of unmodeled effects}
When characterizing a quantum processor we're often interested in how \emph{far} $\cM$ is from a valid model, i.e., ``how much error in the processor does $\cM$ not capture?''.  It may be extremely costly (or entirely prohibitive) to construct a valid model, and a simpler model that fits \emph{most} of the data may be preferable.  As George Box famously said, ``All models are wrong but some are useful'' \cite{BOX1979201}.  To be useful a model doesn't need to capture all of a quantum processor's behavior.

To know when we've captured enough of the behavior, we need to quantify the distance, in some meaningful metric, between the model and the nearest valid model.  If both our model and a known-to-be-valid model are both specified using quantum process matrices, then the diamond norm distance between them would be a good metric.  But usually this isn't an option.  And neither $N_\sigma$ nor $\gamma$ is the right sort of metric --- they quantify the amount of evidence, not the amount of error.  This is clear from the fact that just increasing the amount of data taken (e.g. increasing $N$) can increase both $N_\sigma$ and $\gamma$, even though nothing about the underlying processes or models is changing (cf.~Eqs.~\ref{eq:logl}, \ref{eq:Nsigma} and \ref{eq:gamma}).  

We recently introduced a metric of unmodeled effects called \emph{wildcard error} \cite{blumekohout2020wildcard}, and we deploy it here.  Here's a concise summary of how wildcard error quantifies unmodeled error.  Any $\mathrm{M}_{\vec{\theta}}$ can be augmented by combining it with a \emph{wildcard error model}, which assigns a certain amount of \emph{wildcard error budget} $w$ to each quantum circuit.  It does so by allocating $w_g$ to each gate $g$, and then computing each circuit's wildcard budget by summing $w_g$ over the gates in it.  Wildcard budget relaxes the base model's predictions in a precisely metered way:  if the base model predicts probabilities $\vec{p}$, and the wildcard model assigns $w$, then any $\vec{p}'$ whose \emph{total variation distance} (TVD) to $\vec{p}$ is $\leq w$ is consistent with the relaxed prediction.  Wildcard models are parameterized by \emph{wildcard error rate} vectors $\vec{W} = \{w_g\}$, and augmenting base model $\mathrm{M}_{\vec{\theta}}$ with wildcard error rates $\vec{W}$ yields a \emph{wildcard-augmented model} $\mathrm{M}_{\vec\theta,\vec{W}}$.  To quantify unmodeled error, we find the smallest amount of wildcard error budget (i.e., a minimal $\vec{W}$) that is sufficient to make the wildcard-augmented model consistent with the data.  (If the base model is already consistent, then  $\vec{W}=0$ suffices, indicating there is no unmodeled error).  To determine whether a given $\vec{W}$ is sufficient, we use a standard loglikelihood test, but compute $\mathrm{M}_{\vec\theta,\vec{W}}$'s likelihood by summing up each each circuit's likelihood \emph{maximized} over the TVD-ball of outcome distributions that are consistent with $\mathrm{M}_{\vec\theta,\vec{W}}$'s relaxed predictions \footnote{We perform a loglikelihood test on the full dataset, with a test significance level of $2.5\%$; we also perform a loglikelihood test on the data from each circuit at a significance level of $2.5/N_C \%$ where $N_C$ is the total number of circuits \cite{blumekohout2020wildcard}. This composite test has a family-wise significance of $5\%$.}.  We define a \emph{minimal} wildcard error model as having the smallest $L_1$-norm $\|\vec{W}\|_1$ -- this is a somewhat arbitrary choice, but makes $\|\vec{W}\|_1$ a reasonable measure of total unmodeled error.  Because each $w_g$ represents an amount of TVD per gate, it can be compared directly with standard error metrics with the same ``units'', e.g., diamond norm distance.

We combine wildcard error to a (parameterized) quantum processor model $\cM$ by augmenting the non-parameterized $\mathrm{M}_{\vec{\theta}}$ that achieves the maximum likelihood.  The wildcard-augmented model, $M_{\vec\theta,\vec{W}}$, effectively divides the processor's error processes into two categories: (1) modeled effects, which are captured and predicted by the best-fit model $\mathrm{M}_{\vec\theta}$; and (2) unmodeled effects, which are \emph{not} modeled or predicted at all, but whose impact is upper-bounded by $\vec{W}$.  This division provides as much insight and predictability as possible for effects that the base model \emph{can} explain, while acknowledging and quantifying the total impact of the unmodeled effects.  

If the best-fit model is expressed using process matrices, then the elements of $\vec{W}$ can be compared directly to any TVD-based error metrics (e.g. diamond norm) derived from the base model.  In many situations, this comparison can provide explicit justification for Box's aphorism.  If the magnitude of a gate's unmodeled error ($w_g$) is much smaller than the magnitude of its modeled error, then the base model is ``useful'' because it captures the majority of the error behavior, even if $N_\sigma$ indicates that it is surely ``wrong''.

\subsubsection{Discussion}
$N_\sigma$, $\gamma$, and $\vec{W}$ provide complementary ways of quantifying how well a model fits a dataset.  Their computation is broken down diagrammatically in Fig.~\ref{fig:comparisons}.  $N_\sigma$ and $\gamma$ quantify the evidence of unmodeled effects, whereas $\vec{W}$ measures their size.  Either (or both in concert) can be used to guide the algorithm we present here, and decide whether a given model is ``good enough'' for a given use.  In many scenarios, we expect this choice will depend on whether the experiment's goal is (a) to identify and understand the processor's behavior for scientific reasons, or (b) to determine whether the processor will be able to satisfy engineering requirements.  Statistical weight of evidence ($N_\sigma$, $\gamma$) can identify effects that \emph{exist} whether or not they are important.  Wildcard error can quantify whether effects are likely to be \emph{important} for information-processing tasks.

We adopt the more pragmatic approach, using wildcard error to set our ``good enough'' threshold in Section \ref{sec:example}.

\section{Testing a sequence of nested models\label{sec:method}} 
We now have all the tools and concepts required to give a concrete, precise description of the multi-model characterization technique introduced in Section \ref{sec:multimodel}.  Let $\{\cM_i\}_{i=1}^{m}$ be a sequence of nested models where $\cM_1 \subset \cM_2 \cdots \subset \cM_m$.  We consider here just a 1D chain of nested models, but this could be straightforwardly  generalized into exploration of a tree or lattice of models by considering multiple ``adjacent'' models at each step\footnote{With this 1D chain structure, we only move onto the next, larger model if we reject the null hypothesis that the current model is true. This means that we do not need to increase our $N_{\sigma}$ threshold to account for the fact that we are performing multiple hypothesis tests \cite{Bretz2009-jf}. With a tree or lattice structure, a procedure that, e.g., increases the $N_{\sigma}$ significance threshold to maintain the family-wise error rate of the hypothesis tests is required.}.  Let $k_i$ be the number of parameters of $\cM_i$.  Similarly, let $\{\cC_j\}_{j=1}^l$ be a series of experiment designs (circuit lists).  These can be chosen independently from the models, as models don't technically require any specific or minimal amount of data to be tested.  However, models with more parameters require proportionately more data to estimate all of the parameters accurately, and sometimes a particular experiment design facilitates fitting a model's parameters \cite{nielsen2020gate}.  The experiment designs can also be chosen independently of each other, though we envision the number of circuits in each design and the circuits' size increasing with $j$.  We denote by $\cD_j$ the data from repeating each circuit in $\cC_j$.

The method is iterative.  At each stage it keeps track of a current model and experiment design, which we index using $i$ and $j$ respectively.  At the beginning of each stage, we find $\log\cL(\cM_i, \cD_j)$ by maximizing the log-likelihood over the parameter space.  We assume that $\log\cL(\cM_{ref}, \cD_j)$ is also available, and compute the $2\Delta\log\cL$ and $\Delta k$ values comparing $\cM_i$ to $\cM_{ref}$.  From these, $N_\sigma$ and $\gamma$ (Eqs.~\ref{eq:Nsigma} and \ref{eq:gamma}) are derived, and $\vec{W}$ is computed as described above.  We then decide, based on application-specific criteria involving $N_\sigma$, $\gamma$, and/or $\vec{W}$, whether $\cM_i$ ``sufficiently describes'' $\cD_j$.  This sufficiency condition is an intentionally subjective and flexible criterion, as different applications require different levels of characterization precision.  If the model is sufficient to describe the data, then we move on to the next dataset; if not, we move to the next model.

\begin{algorithm}[H]
\begin{algorithmic}
  \State $\{\cM_i\}_{i=1}^{m} \gets \mbox{sequence of nested models}$
  \State $\{\cC_j\}_{j=1}^{l} \gets \mbox{sequence of circuit lists}$
  \State $i \gets 1$
  \For{$\cC$ in $\cC_1, \cC_2, \ldots \cC_l$}
  \State $\cD \gets \mbox{TakeData}(\cC)$
  \State $s_{ref} \gets 2\log\cL(\cM_{ref}, \cD)$
  \Comment{execute circuits}
  \Repeat
  \State $s \gets 2\log\cL(\cM_i, \cD)$
  \Comment{maximization}
  \State $\Delta s \gets s_{ref} - s$
  \State $\Delta k \gets k_{ref} - k_i$
  \State $N_\sigma \gets \mbox{compute}\,N_\sigma(\Delta s, \Delta k)$
  \Comment{Eq.\ref{eq:Nsigma}}
  \State $\gamma \gets \mbox{compute}\,\gamma(\Delta s, \Delta k)$
  \Comment{Eq.\ref{eq:gamma}}
  \State $\vec{W} \gets \mbox{compute}\,\mbox{$\vec{W}_{\mathrm{min}}$}(\cM_i, \cD, s_{ref}, k_{ref})$
  \Comment{Ref.\cite{blumekohout2020wildcard}}
  \If{$\mbox{ModelIsSufficient}(N_\sigma, \vec{W}, \gamma)$}
  \State \textbf{break}
  \EndIf
  \State $i \gets i + 1$
  \Until{$i > l$}
  \EndFor
\end{algorithmic}
  \caption{Iterative model testing for quantum processor characterization.\label{lst:model_testing_method}}
\end{algorithm}

The entire approach is outlined in Algorithm \ref{lst:model_testing_method}, giving a more concrete summary to the graphical depiction in Fig.~\ref{fig:method_schematic}.  The algorithm consists of an outer loop over experiment designs and an inner loop over models.  The inner loop computes the fit and model selection metrics discussed in Section \ref{sec:testing}.  We use $s$ to hold values of the $2\log\cL$ statistic, and explicitly indicate how the metrics $N_\sigma$ and $\gamma$ only depend on the difference between the compared models' $2\log\cL$ and $k$ values.  Finding a minimal wildcard model independently requires the observed frequency and $\cM_i$'s predicted probability of each of circuit outcome in $\cD$ and so the $\mbox{$\vec{W}_{\mathrm{min}}$}$ function takes $\cM_i$ and $\cD$ as arguments.  The inner loop stops based on the $\mbox{ModelIsSufficient}$ function, which contains customized logic that can be, in general, based on any of the metrics.  When this functions returns \textbf{True} the model is declared to ``sufficiently describe'' the data and the inner loop exits, causing advancement to the next experiment design.  Note that considering a larger dataset does not reset $i$, the model index -- we continue using the final model of the last outer iteration.  The outer loop iterates through successively larger experiment designs and ultimately we either find a model that sufficiently describes $\cD_l$ or we exhaust all our models before this happens.

Algorithm \ref{lst:model_testing_method} is a flexible procedure for quantum processor characterization that can be applied to almost any situation where one or more models of a quantum processor are available.  Its flexibility originates from the freedom to choose the models, experiment designs, and sufficiency criterion it utilizes.  Let us briefly discuss several factors that help inform these choices.

First, it should be noted that the models must be simulated in order to compute fit metrics.  This sets a practical limit on the size of both the models and experiment designs that can be considered.  Another factor that may limit the complexity of the experiment designs is the intended application of the processor.  For instance, if a processor is only expected to ever run one type of circuit, then it may be acceptable to only test circuits of this type.

There are also many reasonable choices of a sufficiency condition (the $\mbox{ModelIsSufficient}$ function in Algorithm \ref{lst:model_testing_method}).  In the worked example of Section \ref{sec:example} we define sufficiency to mean that unmodeled errors are small in size, and base it entirely on the wildcard error rate vector, $\vec{W}$.  Sufficiency could also be implemented as a specific certainty that the model is valid, in which case $\mbox{ModelIsSufficient}$ would impose a threshold on $N_\sigma$.

\section{Application to a 2-qubit processor\label{sec:example}}
In this section we demonstrate the characterization technique described in the previous section by applying it to simulated data.  We consider a 2-qubit processor with x- and y-axis $\pi/2$ rotation gates, $G_x^{(i)}$ and $G_y^{(i)}$, on each qubit $i=0,1$, and two CNOT gates between the qubits, $G_{cnot}^{(0\rightarrow 1)}$ and $G_{cnot}^{(1 \rightarrow 0)}$.

\subsection{The noisy processor}
We add errors to each gate by following the ideal ``target'' gate with an exponentiated \emph{error generator} composed as a linear combination of ``Hamiltonian'' and ``stochastic'' \emph{elementary error generators} $H_P$ and $S_P$, respectively \cite{blumekohout2021taxonomy}.  These elementary generators are indexed by a Pauli $P$ and act on density matrices $\rho$ by
\begin{eqnarray}
  H_P & : & \rho \rightarrow i[P, \rho] \quad \mbox{and} \\
  S_P & : & \rho \rightarrow P\rho P - \rho.
\end{eqnarray}
$H_P$ generates coherent (generalized over-rotation) errors about the $P$-axis, and $S_P$ generates incoherent (generalized dephasing) errors that diminish qubit coherence in the Pauli directions that do not commute with $P$.  If $G_0$ is the Pauli transfer matrix of an ideal $k$-qubit gate, then
\begin{equation}
  G = \exp\left(\sum_P h_P H_P + s_P S_P\right) G_0
\end{equation}
is the Pauli transfer matrix for the corresponding noisy gate of the processor. Here $P$ ranges over all $k$-qubit Pauli matrices and the $h_P$ and $s_P$ coefficients determine the strength of each type of error.  These error types preserve completely-positive trace-preserving (CPTP) maps, and so we are guaranteed that $G$ is a CPTP map.  The errors present in our artificial 2-qubit processor are given in Table \ref{table:datagen_errors}.  The state preparation and measurement (SPAM) errors are similarly constructed by following or preceding the ideal operation by an exponentiated error generator.  The coefficients of these generators are specified in the $\rho$ and $M$ rows of Table \ref{table:datagen_errors}.  The row marked ``all gates'' indicates an additional error that is applied after every gate operation.  The precise magnitudes for these errors are chosen arbitrarily, but their structure is intended to reflect the physically plausible situation where a processor's single qubit gates have over-rotation and dephasing errors about their axis, and there exists a dominant $ZZ$ coupling Hamiltonian that causes errors in the CNOT gates as well as via an always-on background effect.

We generate data from our artificial processor by simulating the outcome counts of each circuit as needed.  We use the noisy gate and SPAM models described by Table \ref{table:datagen_errors} to compute a circuit's outcome probabilities and sample the resulting multinomial probability distribution $N=10000$ times.

\begin{table}
  \begin{tabular}{|c|c|} \hline
    Gate & error generator coefficients \\ \hline
    $G_x^{(0)}$ & $H_X=0.002$, $S_X=0.002$ \\
    $G_y^{(0)}$ & $H_Y=0.002$, $S_Y=0.001$ \\
    $G_x^{(1)}$ & $H_X=0.01$, $S_X=0.0005$ \\
    $G_y^{(1)}$ & $H_Y=0.0015$, $S_Y=0.0001$ \\
    $G_{cnot}^{(0 \rightarrow 1)}$ & $H_{ZZ}=0.06$, $S_{XX}=0.002$ \\
    $G_{cnot}^{(1 \rightarrow 0)}$ & $H_{ZZ}=0.03$, $S_{XX}=0.02$ \\
    $\rho$ & $S_{ZI}=S_{IZ}=0.001$ \\
    $M$ & $S_{ZI}=S_{IZ}=0.001$ \\
    all gates & $H_{ZZ}=0.0002$ \\
    \hline
  \end{tabular}
  \caption{The errors chosen for our artificial 2-qubit quantum processor.  The text explains precisely how these coefficients determine the errors on the gates and SPAM operations of the processors.  The values are chosen to reflect a processor with predominant $ZZ$-type over-rotation errors on its entangling gates, and smaller amounts of over-rotation and dephasing on its single-qubit gates.  A $ZZ$ term is also present, which occurs after every gate (but not SPAM) operation.\label{table:datagen_errors}}
\end{table}

\subsection{An initial benchmark}
\begin{figure}
  \begin{center}
    \includegraphics[width=3.7in]{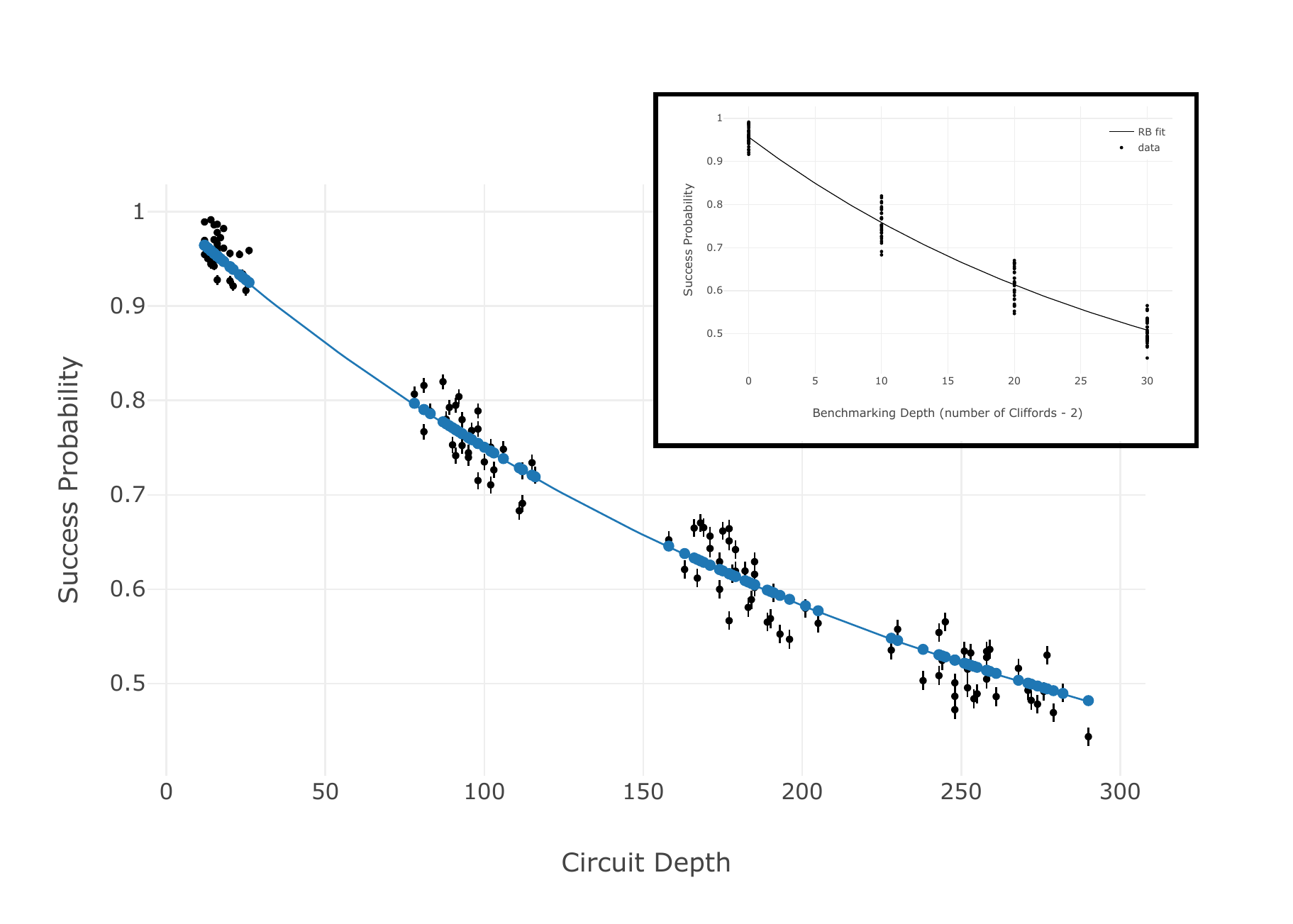}
  \end{center}
  \caption{Simulated data from our artificial 2-qubit processor, for the initial set of circuits $\mathcal{C}_1$.  There are a total of 120 RB circuits (30 at each of 4 Clifford depths), and we plot their success probability as a function of circuit native-gate depth (black points).  This depth is evaluated \emph{after} compiling each random Clifford operation into the set of native gates.  The inset shows the same data plotted against the Clifford depth, and the fit gives an RB number of $\approx 0.029$.  The error bars on the data indicate a 2$\sigma$ standard error (each circuit was repeated $N=10,000$ times).  The blue points and line are the predictions from our initial depolarizing noise model, $\mathcal{M}_1$.\label{fig:rb_data}}
\end{figure}

Let us suppose that we are handed the 2-qubit processor just described, and asked to characterize it.  Knowing nothing about the processor, an intuitive first step would be to run a holistic benchmark on the processor.  Randomized benchmarking is a good choice, so let us run a set of RB circuits.  Since we have just 2 qubits we can use standard Clifford RB here; for more qubits, we would use a scalable variant of RB such as direct RB \cite{Proctor2019-wp} or a different benchmark \cite{Cross2019-ku, Boixo2018-kp,proctor2020measuring}.  We choose 30 random Clifford circuits at Clifford-counts of 2, 12, 22, and 32, for a total of 120 circuits.  The RB data we obtained is plotted as a function of the Clifford depth in the inset of Fig.~\ref{fig:rb_data}.  The computed RB number is $\approx 0.029$.

The RB number is often interpreted as an error rate, and used to define a depolarizing noise model.  RB does not guarantee that this procedure will result in a useful model, and so we will \emph{not} do this, but will instead ask whether \emph{any} depolarizing noise model is able to fit the RB data.  Separate from the RB analysis we construct a depolarizing noise model and fit it to the RB data.  The best-fit model's predicted success probabilities are show in Fig.~\ref{fig:rb_data} by the blue line and points.  The data are plotted as a function of native gate depth to spread the points horizontally and to show how the y-direction scatter at a Clifford depth is partially explained by the varied gate depth.  The $2\sigma$ standard error bars plotted on the data (black) points reveal that the model does not explain the data to the expected statistical precision.

\subsection{Applying the multi-model approach}
This motivates the consideration of richer models, and so we now apply the characterization method of Section \ref{sec:method}.  We would like to know, in the end, how our processor behaves on general circuits.  Since we've already taken RB data, it is natural to set $\mathcal{C}_1$ as our set of 120 RB circuits.  RB circuits are designed to be insensitive to coherent noise and so we also include a set of periodic circuits designed to be sensitive to coherent errors.  We choose a set of GST-like circuits, each composed of a repeated germ sub-circuit sandwiched between two fiducial sub-circuits (cf.~\onlinecite{nielsen2020gate}).  We find a set of 436 circuits that are sensitive to all (Markovian) coherent errors and have up to 16 germ repetitions.  These are added to the 120 RB circuits to form $\mathcal{C}_2$.  We note in passing that these 556 circuits are far fewer than the more than $9,000$ circuits required to perform standard GST.  GST requires a much larger number because it uses an over-complete set of circuits and its standard set of circuits amplifies \emph{every} Markovian gate error.

We next decide on our sequence of models, $\{\cM_i\}_{i=1}^{m}$.  For this example, we consider the following nested models:
\begin{enumerate}
\item \textbf{Depolarizing model $\cM_1$}: each gate has the same depolarization rate.  The state preparation and measurement are also depolarized, each at an independent rate.  This model, then, has a total of $3$ parameters.
\item \textbf{Gate-dependent depolarizing model $\cM_2$}:  each gate has an independent depolarization rate, and the state preparation and measurement have independent rates for each qubit, giving the model $10$ parameters.
\item \textbf{Pauli-stochastic model $\cM_3$}:  each gate now has independent stochastic error rates along each Pauli direction.  Each single qubit gate is given by 3 error rates and each two-qubit gate by 15 error rates.  State preparation and measurement (SPAM) operations are allowed only local errors, and so have 3 degrees of freedom per qubit (6 total). This brings the total number of parameters to $4 \times 3 + 2 \times 15 + 6 + 6 = 54$.
\item \textbf{Hamiltonian + Pauli-stochastic model $\cM_4$}:  the same as $\cM_3$ except each operation also is allowed Hamiltonian (i.e. over-rotation) errors along each Pauli axis, doubling the number of parameters to $108$.
\item \textbf{Full CPTP model $\cM_5$}:  each gate is allowed to be an arbitrary CPTP map, and SPAM operations are allowed to be followed or preceded (respectively) by such a map.  The total number of parameters for this model is $2160$.  This is the model that standard GST uses from the outset.
\end{enumerate}

The freedom to chose any set of nested models gives the method great flexibility.  In our case, we presumed to know nothing about the types of noise that might appear and chose a series of models that capture generic types of noise and that are not specifically tailored to our processor.  If, for example, we had reason to expect that $ZZ$-type over-rotation errors would be dominant, we could have included a model with only these types of entangling errors.

\begin{table}
  \begin{tabular}{|c|c|c|c|c|c|c|} \hline
    Iteration & Circuits & Model & $N_\sigma$ & $\gamma$ & $\max(\vec{W})$ \\ \hline
    1 & $\cC_1$ & $\cM_1$ & 653  & 50.0 & 0.0029 \\
    2 & $\cC_1$ & $\cM_2$ & 529  & 41.0 & 0.0021 \\
    3 & $\cC_1$ & $\cM_3$ & 13   & 2.1  & 0.00034 \\
    4 & $\cC_2$ & $\cM_3$ & 641  & 23.6 & 0.026  \\
    5 & $\cC_2$ & $\cM_4$ & -0.5 & 1.0  & 0       \\
    \hline
  \end{tabular}
  \quad\raisebox{-0.5in}{\includegraphics[width=0.7in]{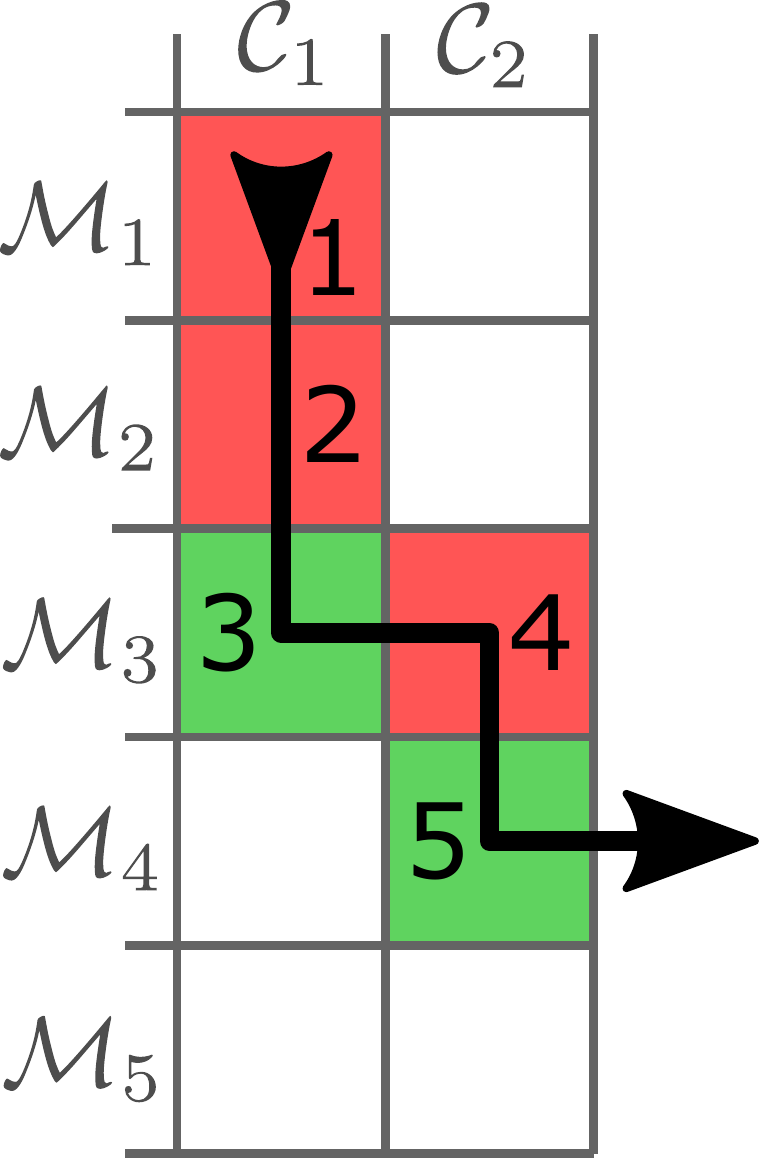}}
  \caption{The primary outputs from running Algorithm \ref{lst:model_testing_method} on a simulated 2-qubit processor. Each row is an iteration of the algorithm, where we compare the model (third column) to the data generated by a set of circuits (second column).  $N_\sigma$, $\gamma$, and wildcard error rate $\vec{W}$ values are computed, all with respect to a maximal model $\cM_{ref}$ as described in the text.  The decision of whether a model sufficiently describes the data, given by Eq.~\ref{eq:stopping_criterion}, only depends on the maximum element of $\vec{W}$, so that is all that is included here ($\vec{W}$ in its entirety is given in Fig.~\ref{fig:example_results}).  The diagram to the right shows the numbered iterations in the format of Fig.~\ref{fig:method_schematic}\label{table:example_results}}
\end{table}

The last necessary ingredient is a model acceptance criterion, i.e., the $\mbox{ModelIsSufficient}$ function in Algorithm \ref{lst:model_testing_method}.  We only demand that a model describe most of the behavior of the processor, and not that the model be valid from a statistical standpoint.  Specifically, we define our criterion as a simple $\epsilon = 10^{-3}$ threshold on the maximum element of $\vec{W}$.  That is,
\begin{equation}
  \mbox{ModelIsSufficient}(N_\sigma, \vec{W}, \gamma) \equiv \max(\vec{W}) < \epsilon.\label{eq:stopping_criterion}
\end{equation}
This criterion implies that a model is satisfactory when the worst gate's unmodeled TVD allocation is less than $0.1\%$.  The choice of $\epsilon$ here is somewhat arbitrarily, and different applications may be more or less willing to tolerate unmodeled error.

Execution of Algorithm \ref{lst:model_testing_method}, given our inputs, produces the results of Table \ref{table:example_results}.  The table shows $N_\sigma$ and $\vec{W}$ at each iteration of the characterization process.  The algorithm begins by rejecting the depolarizing and then the gate-dependent depolarizing models based on the RB data ($\cD_1$). Model $\cM_3$, which allows independent Pauli stochastic errors, is tested next, and is able to describe the RB data well enough to be accepted, and causes the algorithm to advance to $\cD_2$.  The periodic data of $\cD_2$ causes $\cM_3$ to be rejected, leading to a test of the 108-parameter $\cM_4$, which allows coherent and Pauli-oriented stochastic errors.  $\cM_4$ is unsurprisingly capable of modeling the data very well, as the model we used to generate the data only included these types of errors. $\cM_4$ is accepted.  Since $\cC_2$ is our final experiment design the algorithm exits without considering $\cM_5$.

\begin{figure}[h]
  \begin{center}
    \includegraphics[width=3.5in]{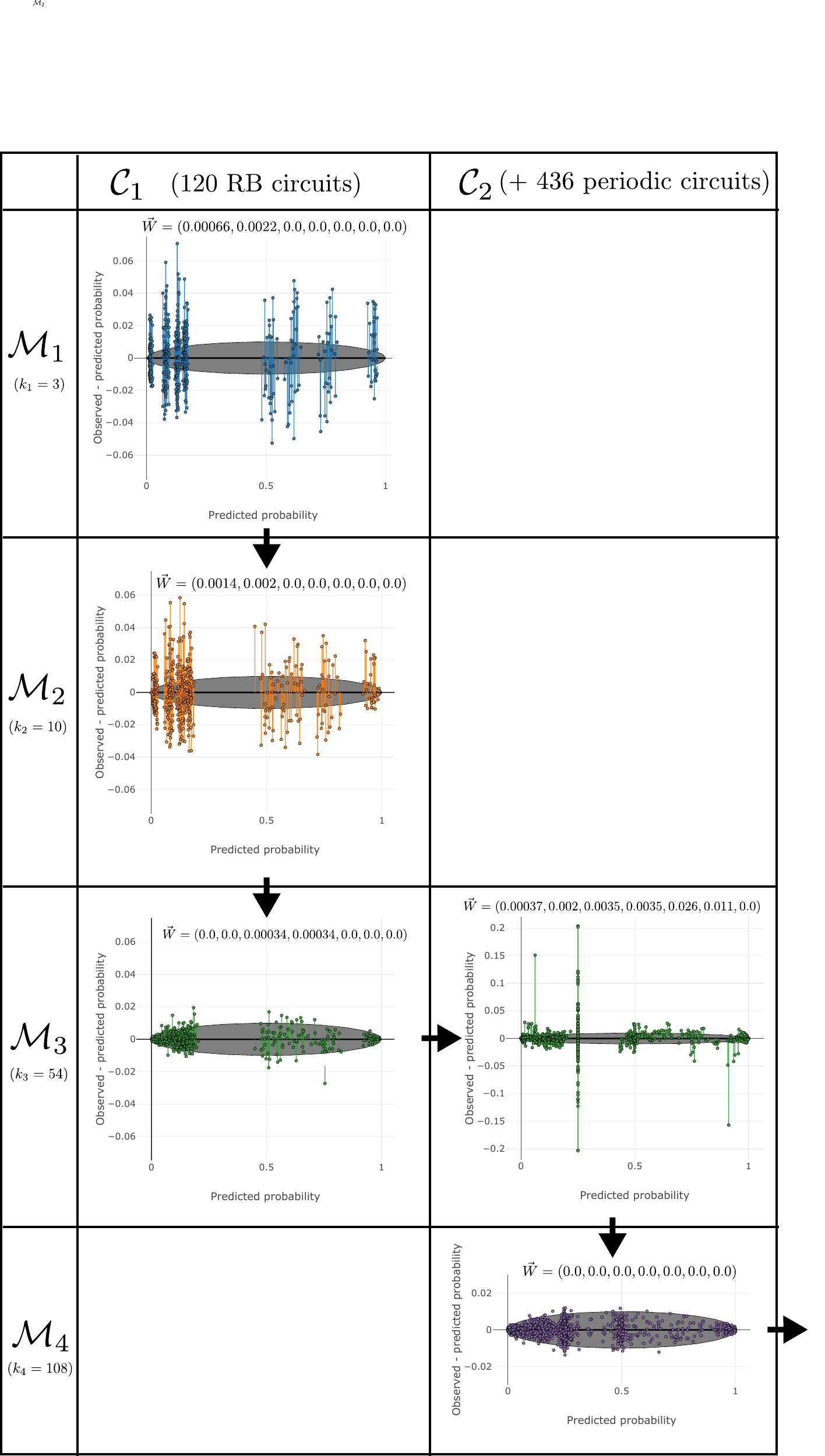}
  \end{center}
  \caption{Differences between observed frequencies and predicted probabilities at each algorithm step.  Plots are arranged in a grid pattern similar to that of Fig.~\ref{fig:method_schematic} and show, as a function of the predicted probabilities, $p$, the difference between this probability and the observed frequency $f$.  In each plot, there is one point per circuit outcome.  Lines drawn from each point toward the x-axis show how the given wildcard budget $\vec{W}$ is able to adjust $p$ toward $f$ ($f-p$ toward $0$).  Shaded regions indicate $2\sigma$ standard error bars.\label{fig:example_results}}
\end{figure}

To visualize how well each model is able to reproduce the datasets to which it was fit to, Fig.~\ref{fig:example_results} plots the difference between the observed frequency (a fraction of the total counts) and predicted probability for every circuit outcome in the dataset.  Differences are plotted as a function of the predicted probability, and shaded regions demarcate a $2\sigma$ standard error bar (i.e., where $|p-f| < 2\sqrt{\frac{p(1-p)}{N}}$ when $f$ and $p$ are the frequency and probability respectively).  For a statistically valid model, we expect $\approx 95\%$ of the points to lie within the shaded region.  When they do not, a wildcard model must account for the remaining discrepancy and $|\vec{W}| > 0$.  Vertical lines emanating from the points indicate how far that point is allowed to move toward the x-axis given the minimal wildcard error rates $\vec{W}$, that were needed.  

As a point of reference, the distances between blue and black points in Fig.~\ref{fig:rb_data} are the successful-outcome subset of the all the points in the first frame of Fig.~\ref{fig:example_results}. The error bars on Fig.~\ref{fig:rb_data}'s points correspond to the height of shaded region in Fig.~\ref{fig:example_results}.  We see from Fig.~\ref{fig:example_results} how more complex models are able to better predict the data, and how smaller wildcard error models are needed to augment such models.

This simple example illustrates several noteworthy points.  
\begin{enumerate}
\item It is clear that $N_\sigma$ provides different information from $\vec{W}$.  The first two iterations have $N_\sigma > 500$, expressing certainty that these models do not describe all the data. But the corresponding wildcard models have maximum elements less than $3\times 10^{-3}$, indicating that if just this small amount of additional (``wildcard'') TVD is allowed per gate, the data can be explained by the model.  Indeed, if we had set $\epsilon = 10^{-2}$, then we would have accepted the depolarizing model ($\cM_1$) and immediately moved to $\cD_2$ for the second iteration.  In the third iteration, we treat an ostensibly invalid model ($N_\sigma = 13$) as sufficiently describing $\cD_1$ based on the wildcard error $\vec{W}$ being small.  In the final iteration we find that $\cM_4$ is a valid model and so necessarily doesn't require any wildcard error.
\item Different circuit lists are sensitive to different types of errors.  These results show, unsurprisingly, that RB circuits are insensitive to, and effectively mask, coherent errors.  Even though the underlying processor possesses predominantly coherent errors (cf.~Table \ref{table:datagen_errors}) the purely stochastic model $\cM_3$ is able to fit the RB data very well (and would be considered a valid model if only 1,000 samples were taken for each circuit).  Indeed, if we only considered $\cC_1$ it would be extremely difficult or impossible to determine whether the errors were coherent or incoherent using only RB data.
\item Finally, this example illustrates how repeated model testing makes efficient use of experimental data.  We note that through the third iteration only the initial set of RB data was utilized.  By this point in the algorithm we have constructed a 54-parameter model of the stochastic errors that clearly is able to explain more of the data than a simple depolarizing model (cf.~Fig.~\ref{fig:example_results}).  This stands in stark contrast to the single average-error-per-Clifford number obtained by a standard analysis of the \emph{same} data.  It is also the case that all the models up to this point ($\cM_1$ to $\cM_3$) can be simulated efficiently, and so are easily scalable to 10s or even 100s of qubits.
\end{enumerate}

\section{Conclusions}
Iterative model testing is a simple, powerful technique for learning about the behavior of quantum processors.  We have shown how the techniques from statistics, along with the novel concept of wildcard error, can be applied to a series of nested quantum processor models to determine a good model.  We have outlined a general procedure for performing such testing, and have demonstrated its utility by characterizing an simulated 2-qubit processor.  By being flexible with regard to the models that are tested and the data that is utilized, our method can be applied over a wide range of scenarios and adapted to define what constitutes a ``good'' model based on a processor's intended application.  The presented algorithm considers increasingly rich datasets, and tests sequentially more complex models.  This allows it to avoid expending resources until they are absolutely needed, improving upon existing techniques such as gate set tomography.  Throughout the characterization process, we quantify how much of the processor's error is not being captured by the current model.  In the end, either an acceptable model is found or we have attempted the most complex model available (or feasible) to us.  In either case, the amount of unmodeled error gives us a concrete sense of how the model can be used, and ensures that it never fails to provide useful information.  We find, overall, that this model-testing approach yields more detailed characterization information using the same experimental resources (data) when compared with existing techniques.


This work was supported by the U.S. Department of Energy, Office of Science, Office of Advanced Scientific Computing Research, and the Laboratory Directed Research and Development program at Sandia National Laboratories. Sandia National Laboratories is a multi-program laboratory managed and operated by National Technology and Engineering Solutions of Sandia, LLC., a wholly owned subsidiary of Honeywell International, Inc., for the U.S. Department of Energy’s National Nuclear Security Administration under contract DE-NA-0003525. All statements of fact, opinion or conclusions contained herein are those of the authors and should not be construed as representing the official views or policies of the U.S. Department of Energy, or the U.S. Government.

\bibliography{citations}

\end{document}